# Lexicodes over Rings

Kenza Guenda, T. Aaron Gulliver and S. Arash Sheikholeslam *

April 26, 2018

**Abstract**

In this paper, we consider the construction of linear lexicodes over finite chain rings by using a $B$-ordering over these rings and a selection criterion. As examples we give lexicodes over $\mathbb{Z}_4$ and $\mathbb{F}_2 + u\mathbb{F}_2$. It is shown that this construction produces many optimal codes over rings and also good binary codes. Some of these codes meet the Gilbert bound. We also obtain optimal self-dual codes, in particular the octacode.

*K. Guenda is with the Faculty of Mathematics USTHB, University of Science and Technology of Algiers, Algeria. T. A. Gulliver and S. A. Sheikholeslam are with the Department of Electrical and Computer Engineering, University of Victoria, PO Box 3055, STN CSC, Victoria, BC, Canada V8W 3P6. email: agullive@ece.uvic.ca.

# 1 Introduction

Surprisingly, many good binary linear codes can be constructed using the following greedy algorithm with minimum distance as the selection criterion. Starting with the all zero vector, all binary vectors of length $n$ are considered in lexicographic order, and when the distance of a vector to all other vectors in the code is at least $\delta$, the vector is added to the code. Levenstein [20] proved that the resulting code (called a lexicode), is linear. Conway and Sloane [9] proved that the lexicodes are linear over fields of order $2^{2^l}, l \in \mathbb{N}$. Moreover, they proved linearity when using a more general selection criterion called a turning-set.

Brualdi and Pless [8] presented another generalization of binary lexicodes. They introduced the concept of a $B$-ordering, which is used in the greedy algorithm instead of the standard basis. Their starting point is a list of binary vectors of length $n$, ordered lexicographically with respect to a basis obtained by adding recursively all previous words to the next basis word. They proved that the resulting lexicodes are also linear. Unfortunately, for fields other than $\mathbb{F}_2$, the lexicodes constructed using a $B$-ordering are not always linear. To solve this problem, Bonn [7] introduced another concept called forcing linearity. In this case, a list of all vectors over $\mathbb{F}_q$ of length $n$ is searched. This list need not be ordered in a specific way. If a vector $\mathsf{a}$ satisfying $d(\mathsf{a}, \mathsf{y}) \geq \delta$ is found, then $\mathsf{a}$ is added to the lexicode as well as all its multiples without checking the minimum distance condition. Surprisingly, this condition is satisfied for all added words [7, Proposition 1]. Thus the resulting code, which is forced to be linear over all finite fields, has a basis composed of the selected vectors $\mathsf{a}$ and has minimum distance greater than or equal to the designed distance $\delta$.

Recently, van Zanten and Nengah Suparta [23, 24] generalized the work of Bonn to a more general selection property over an arbitrary finite field $\mathbb{F}_q$. They considered a $B$-ordering on $\mathbb{F}_q^n$. By using a multiplicative selection property $P$, they proved that the resulting lexicode $C(B, P)$ is linear and each vector $\mathsf{x} \in C(B, P)$ satisfies the property $P$.

In this paper, the construction of lexicodes is considered using a $B$-ordering over finite chain rings and a selection criterion. First, the concept of a $B$-ordering is generalized to finite chain rings. Then we consider the selection property. A greedy algorithm over finite chain ring is given which is based on these results. As examples, we consider the rings $\mathbb{Z}_4$ and $\mathbb{F}_2 + u\mathbb{F}_2$, and construct lexicodes using different bases and different properties. As a special case, greedy algorithms are given to find self-orthogonal codes. In particular, the octacode $O_8$ is obtained as a lexicode over $\mathbb{Z}_4$ using the minimum distance criterion. In this case we also prove that the corresponding binary image meets the Gilbert bound. Tables of lexicodes over the rings $\mathbb{Z}_4$ and $\mathbb{F}_2 + u\mathbb{F}_2$ are given which have been constructed using several selection criterion. We compare the codes obtained over $\mathbb{Z}_4$ with the best codes in [2] and [12].

## 2 Preliminaries

For codes over finite chain rings one can refer to [16]. A finite chain ring $R$ is a local principal ideal ring with maximal ideal $\mathfrak{m} = \langle \gamma \rangle$, where $\gamma$ is a nilpotent element of $R$ with nilpotency index $e$. Hence the elements of $R \setminus \langle \gamma \rangle = R^*$ are units and the ideals of $R$ form the following chain

$$\langle 0 \rangle = \langle \gamma^e \rangle \subsetneq \langle \gamma^{e-1} \rangle \subsetneq \ldots \subsetneq \langle \gamma \rangle \subsetneq R.$$

Let $\mathbb{F}_{p^r}$ denote the field $R/\langle \gamma \rangle$. Hence we have

$$|R| = p^{re} \tag{1}$$

For an integer $n > 0$, $R^n$ is an $R$-module. A non empty subset of $R^n$ is said to be a linear code over $R$ of length $n$ if it is a submodule of $R^n$.

We denote by $\mathbb{Z}_4$ the commutative ring with elements $\{0, 1, 2, 3\}$ and addition and multiplication modulo 4. It is a finite chain ring with maximal ideal $\langle 2 \rangle$ and nilpotency index 2. The ring $\mathbb{F}_2 + u\mathbb{F}_2$ with $u^2 = 0$ is a finite chain ring with maximal ideal $\langle u \rangle$ and nilpotency index 2. The elements of $\mathbb{F}_2 + u\mathbb{F}_2$ are $\{0, 1, u, \overline{u} = 1 + u\}$ and its residue field is $\mathbb{F}_2$. Multiplication coincides with that of $\mathbb{Z}_4$, while addition coincides with that of $\mathbb{F}_4 = \{0, 1, w, w^2 = w + 1\}$, where $w$ and $w^2$ are replaced by $u$ and $\overline{u}$, respectively. For $\mathsf{x} \in \mathbb{Z}_4^n$, denote the number of components of $\mathsf{x}$ equal to $a$ by $n_a(\mathsf{x})$. Then the Hamming weight of $\mathsf{x}$ is $wt_H(\mathsf{x}) = n_1(\mathsf{x}) + n_2(\mathsf{x}) + n_3(\mathsf{x})$.

The Lee weight of $\mathsf{x}$ is $wt_L(\mathsf{x}) = n_1(\mathsf{x}) + 2n_2(\mathsf{x}) + n_3(\mathsf{x})$, and the Euclidean weight of $\mathsf{x}$ is $wt_E(\mathsf{x}) = n_1(\mathsf{x}) + 4n_2(\mathsf{x}) + n_3(\mathsf{x})$. For $\mathsf{x} \in (\mathbb{F}_2 + u\mathbb{F}_2)^n$, denote the number of components of $\mathsf{x}$ equal to $a$ by $n_a(\mathsf{x})$. Then the Hamming weight of $\mathsf{x}$ is $wt_H(\mathsf{x}) = n_1(\mathsf{x}) + n_u(\mathsf{x}) + n_{\overline{u}}(\mathsf{x})$. The Lee weight of $\mathsf{x}$ is $wt_L(\mathsf{x}) = n_1(\mathsf{x}) + 2n_u(\mathsf{x}) + n_{\overline{u}}(\mathsf{x})$, and the Euclidean weight of $\mathsf{x}$ is $wt_E(\mathsf{x}) = n_1(\mathsf{x}) + 4n_u(\mathsf{x}) + n_{\overline{u}}(\mathsf{x})$. The Hamming, Lee and Euclidean distances $d_H(\mathsf{x}, \mathsf{y})$, $d_L(\mathsf{x}, \mathsf{y})$, $d_E(\mathsf{x}, \mathsf{y})$ between two vectors $\mathsf{x}$ and $\mathsf{y}$ are $wt_H(\mathsf{x}-\mathsf{y})$, $wt_L(\mathsf{x} - \mathsf{y})$ and $wt_E(\mathsf{x} - \mathsf{y})$, respectively. The minimum Hamming, Lee and Euclidean weights, $d_H$, $d_L$ and $d_E$ of $C$ are the smallest Hamming, Lee and Euclidean weights among all nonzero codewords of $C$. A more general definition of weights over rings can be defined as follows.

**Definition 1** *Let $R$ be a finite chain ring. A weight $w_h$ on $R$ is called homogeneous if it satisfies the following assertions:*

*(i)* $\forall x \in C$, $\forall u \in R^* : w_h(x) = w_h(ux)$.

*(ii) There exists a constant $\xi = \xi(w_h) \in \mathbb{R}$ such that*

$$\sum w_h(x)_{x\in U} = \xi|U|,$$

*where $U$ is any subcode of $C$.*

Honold and Nechaev [18] proved that for any finite chain ring there exists a homogeneous weight. Note that the Lee weights defined above for $\mathbb{Z}_4$ and $\mathbb{F}_2 + u\mathbb{F}_2$ are homogeneous weights.

Any linear code over $\mathbb{Z}_4$ or $\mathbb{F}_2+u\mathbb{F}_2$ has a generator matrix of the following form

$$G = \begin{pmatrix} I_{k_1} & A & B_1+\gamma B_2 \\ O & \gamma I_{k_2} & \gamma M \end{pmatrix}$$

where $\gamma = 2$ for codes over $\mathbb{Z}_4$ and $\gamma = u$ for codes over $\mathbb{F}_2+u\mathbb{F}_2$. The matrices $A, B_1, B_2$ and $C$ have entries from $\mathbb{F}_2$, and $O$ is a $k_2 \times k_1$ zero matrix. The code $C$ is said to be of type $4^{k_1}2^{k_2}$.

# 3 Construction of Lexicodes over Finite Chain Rings

Let $R = \{\alpha_1, \ldots, \alpha_m\}$ be a finite chain ring with nilpotency index $e$ and residue field $\mathbb{F}_{p^r}$. Hence from (1) we have $|R| = p^{re} = m$. The free module $R^n$ is a linear code over $R$ with basis $B = \{\mathsf{b}_1, \ldots, \mathsf{b}_n\}$. With respect to this basis we give the following $B$-ordering. Recursively define the lexicographically ordered list $V_i = \mathsf{x}_1, \mathsf{x}_2, \ldots, \mathsf{x}_{p^{rei}}$ as follows

$$V_0 := 0,$$

$$V_i := V_{i-1}, \quad \alpha_1\mathsf{b}_i + V_{i-1}, \alpha_2\mathsf{b}_i + V_{i-1}, \ldots, \alpha_m\mathsf{b}_i + V_{i-1}, 1 \leq i \leq n.$$

In this way $|V_i| = m^i$, and $R^n$ is given by $V_n$. Assume now that we have a property $P$ which can test if a vector $\mathsf{c} \in R$ is selected or not. Recall that a selection property $P$ on $V$ can be seen as a boolean valued function $P : V \longrightarrow \{\text{True, False}\}$ that depends on one variable. Over a finite chain ring $R$, the property $P$ is called a multiplicative property if $P[\mathsf{x}]$ is true implies $P[\beta\mathsf{x}]$ is true for all $\beta \in R^*$. As mentioned in the introduction, the first lexicodes were obtained using a weight criterion over finite fields. We now prove that the weight criterion is a multiplicative property over finite chain rings when considering a homogenous weight.

**Lemma 2** *Let $R$ be a finite chain ring, $\delta$ a positive integer and $w_h$ a homogeneous weight on $R$. The property $P[\mathsf{x}]$ is true if and only if $w_h(\mathsf{x}) \geq \delta$ is a multiplicative property.*

**Proof.** We need to prove that if $w_h(\mathsf{x}) \geq \delta$, then $w_h(\beta\mathsf{x}) \geq \delta$ for all $\beta \in R^*$. This is always true from Definition 1 (i). Hence the result follows. ■

Assume now that we have a selection property which is multiplicative over a finite chain ring $R$. The following greedy algorithm provides lexicodes over $R^n$.

**Algorithm A**

1. $C_0 := 0; i := 1;$

2. select the first vector $\mathsf{a_i} \in V_i \setminus V_{i-1}$ such that $P[\gamma^j \mathsf{a_i} + \mathsf{c}]$ for all $0 \leq j \leq e-1$ and all $\mathsf{c} \in C_{i-1}$;

3. if such an $\mathsf{a}_i$ exists, then $C_i := C_{i-1}, \alpha_1\mathsf{a}_i + C_{i-1}, \alpha_2\mathsf{a}_i + C_{i-1}, \ldots, \alpha_m\mathsf{a}_i + C_{i-1}$; otherwise $C_i := C_{i-1}$;

4. $i := i + 1$; return to 2.

For $0 < i \leq n$, the codes $C_i$ are forced to be linear because we take all linear combinations of the selected vectors $\mathsf{a}_{i1}, \ldots, \mathsf{a}_{il}$; $l \leq i$. The codes $C_i$ have a generating set formed by the selected vectors $\mathsf{a}_{i1}, \ldots, \mathsf{a}_{il}$.

Considering the greedy algorithm [24, Algorithm A] for finite fields, a natural question that arises is, can a vector $\mathsf{x} \in V_i \setminus V_{i-1}$ exist with $P[\mathsf{x} + \mathsf{c}]$ for all $\mathsf{c} \in C_i$ and $\mathsf{x} \notin C_i$ ? The following lemma, which is an extension of [24, Theorem 2.1], shows that such a vector does not exist.

**Lemma 3** *Let $R$ be a finite chain ring with maximal ideal $\langle \gamma \rangle$ and nilpotency index $e$. Let $P$ be a multiplicative property over $R$, and let $\mathsf{a_i} \in V_i$ be such that $P[\gamma^j \mathsf{a_i} + \mathsf{c}]$ for all $0 \leq j \leq e-1$ and for all $\mathsf{c} \in C_{i-1}$, $i \geq 1$. Then every $\mathsf{x} \in V_i \setminus V_{i-1}$ satisfying $P[\gamma^j \mathsf{x} + \mathsf{c}]$, for all $0 \leq j \leq e-1$ and all $\mathsf{c} \in C_i$, is in $C_i$.*

**Proof.** The proof is by induction on $i$. Let $l > 0$ be the first index such that $P[\gamma^j \mathsf{a}_l]$ for all $0 \leq j \leq e-1$. Hence $C_0 = C_1 = \ldots = C_{l-1} = \{0\}$, $C_l = \{0, \alpha_1 \mathsf{a}_l, \ldots, \alpha_m \mathsf{a}_l\}$. Let $\mathsf{x} \in V_l \setminus V_{l-1}$ be a vector such that $P[\gamma^j \mathsf{x} + \alpha \mathsf{a}_l]$, for $0 \leq j \leq e-1$ and $\alpha \in \{\alpha_1, \ldots, \alpha_m\}$. Since $\mathsf{x} \in V_l \setminus V_{l-1}$, we can write $\mathsf{x} = \beta \mathsf{a_l} + \mathsf{v}$ for some $\beta \neq 0$ and some $\mathsf{v} \in V_{l-1}$. If $\mathsf{v} = 0$, then we have $\mathsf{x} = \beta \mathsf{a}_l$, and hence $\mathsf{x} \in C_l$. If $\mathsf{v} \neq 0$, for $0 \leq j \leq e-1$ take recursively $\alpha = -\gamma^j \beta$, which gives $P[\gamma^j \mathsf{v}]$, for $0 \leq j \leq e-1$. This contradicts the assumption on $l$.

Let $\mathsf{a_i} \in V_i$, $i > l$, be a selected vector such that $P[\gamma^j \mathsf{a_i} + \mathsf{c}]$ for all $0 \leq j \leq e-1$ and all $\mathsf{c} \in C_{i-1}$. Assume that the lemma holds for all relevant index values less that $i$. Now let $\mathsf{x} \in V_i \setminus V_{i-1}$ be such that $P[\gamma^j \mathsf{x} + \mathsf{c}]$ for all $0 \leq j \leq e-1$ and all $\mathsf{c} \in C_i$. Since $\mathsf{x} \in V_i \setminus V_{i-1}$, we can write $\mathsf{x} = \beta \mathsf{a}_i + \mathsf{v}$ for some $\mathsf{v} \in V_{i-1}$ and $\beta \neq 0$. If we take $\mathsf{c} = -\gamma^j \beta \mathsf{a}_i + \mathsf{c}'$, it follows that $P[\gamma^j \mathsf{v} + \mathsf{c}']$ for all $\mathsf{c}' \in C_{i-1}$ and $0 \leq j \leq e-1$. From the induction assumption we have that $\mathsf{v} \in C_{i-1}$. Since $\mathsf{x} = \beta \mathsf{a_i} + \mathsf{v}$, it must be that $\mathsf{x} \in C_i$. ■

Lemma 3 shows that when a vector $\mathsf{a_i} \in V_i$ is found in Step 2 of Algorithm A, and after extending the list of codewords in Step 3, we can continue the selection procedure by searching the sublist $V_{i+1} \setminus V_i$. Thus at the end of Algorithm A we have a nested sequence of linear codes

$$0 = C_0 \subseteq C_1 \subseteq \ldots \subseteq C_n.$$

The set $B = \{\mathsf{a}_{i1}, \ldots, \mathsf{a}_{il}\}$ is a generating set for the code $C_i$. The code $C_n$ is the lexicode and since it depends only on the selection property $P$ and the ordering $B$, we denote $C_n$ by $C(B, P)$. The lexicode $C(B, P)$ is a maximal code in the sense that it cannot be contained in a larger code with the same generating set and the same property.

**Remark 1** *Our definition of the multiplicative property differs from that of van Zanten and Nengah Suparta [24]. They defined a multiplicative property over a finite field as a boolean valued function $P$ for which $P[\mathsf{x}]$ implies $P[\alpha \mathsf{x}]$ for all $\alpha \in \mathbb{F}_q$. Since $P[\mathsf{a}_i + c]$ holds, then $P[\mathsf{a}_i + \alpha c]$ from Step 2 of [24, Algorithm A]. If the property $P$ is multiplicative, then $P[\alpha^{-1}(\mathsf{a}_i + \alpha c)] = P[\alpha^{-1}\mathsf{a}_i + c]$ for all $\alpha \in F_q$. This is no longer true over rings since there are* *zero divisors. Hence there are some vectors* $\mathsf{c} \in C_i$ *which are missing and may not satisfy the property* $P$*, even if the code is linear and the property is multiplicative. This justifies our modification of the multiplicative property and adding the constraint in Step 2 to satisfy* $P[\gamma^j \mathsf{a}_i + c]$*, for all* $0 \leq j \leq e-1$*.*

Now we extend [24, Theorem 2.2] to lexicodes over $R$.

**Theorem 4** *For any basis* $B$ *of* $R^n$ *and any multiplicative selection criterion* $P$*, the lexicode* $C(B, P)$ *is linear and* $P[\mathsf{x}]$ *holds for each codeword* $\mathsf{x} \neq 0$*.*

**Proof.** The linearity of the code is assured by the code construction. Since $P[\gamma^j \mathsf{a_i} + \mathsf{c}]$ for all $0 \leq j \leq e-1$ and all $\mathsf{c} \in C_{i-1}$, and the property $P$ is multiplicative, then for all $\mathsf{c} \in C_{i-1}$, we also have $P[\beta\gamma^j \mathsf{a_i} + \beta \mathsf{c}]$, for all $\beta \in R^*$ and $0 \leq j \leq e-1$. Since $C_{i-1}$ is linear, this is equivalent to $P[\beta\gamma^j \mathsf{a_i} + \mathsf{c}]$ for all $\mathsf{c} \in C_{i-1}$. Applying this result for $i = 1, 2, \ldots, k$ sequentially yields that $P[\mathsf{x}]$ is true for any codeword $\mathsf{x} \neq 0$, since the vectors $\mathsf{a_1}, \mathsf{a_2}, \ldots, \mathsf{a_k}$ constitute a generating set for the code $C(B, P)$. ■

# 4 Lexicodes over $\mathbb{Z}_4$

The ring $\mathbb{Z}_4$ is a finite chain ring with maximal ideal $\langle 2 \rangle$ and nilpotency index 2. In this section, we present constructions of lexicodes using different selection properties. We begin with self-orthogonal codes.

## 4.1 Self-Orthogonal Codes

Let $\mathsf{x} = x_1 \ldots x_n$ and $\mathsf{y} = y_1 \ldots y_n$ be two elements of $\mathbb{Z}_4^n$. The inner product of $\mathsf{x}$ and $\mathsf{y}$ in $\mathbb{Z}_4^n$ is defined as $\mathsf{x} \cdot \mathsf{y} = x_1y_1 + \ldots + x_ny_n \pmod 4$. Let $C$ be a $\mathbb{Z}_4$ linear code of length $n$. The dual code of $C$ is defined as $C^\perp = \{\mathsf{x} \in \mathbb{Z}_4 | \ \mathsf{x} \cdot \mathsf{c} = 0 \text{ for all } \mathsf{c} \in C\}$. A code is said to be self-orthogonal if $C \subset C^\perp$.

Consider now the property $P[\mathsf{x}]$ is true if and only if $\mathsf{x} \cdot \mathsf{x} = 0$. This is a multiplicative property over $\mathbb{Z}_4$ because $3\mathsf{x} \cdot 3\mathsf{x} = \mathsf{x} \cdot \mathsf{x} = 0$. Using Algorithm A and this selection property, we produce a linear lexicode $C(B, P)$ over $\mathbb{Z}_4^n$. Hence from Theorem 4 we have that the code $C(B, P)$ for this criterion is linear and $P[\mathsf{x}]$ holds for all $\mathsf{x} \in C(B, P)$. Several lexicodes over $\mathbb{Z}_4^n$ obtained using the selection property $\mathsf{x} \cdot \mathsf{x} = 0$ are given in Table 1. In the case of lexicodes over fields, this selection criterion is sufficient to produce self-orthogonal codes. However this is not the case with $\mathbb{Z}_4$, since the argument of [24, Corollary 5.1] is not true over rings, namely we can have $\mathsf{x} \cdot \mathsf{x} = 0$, $\mathsf{y} \cdot \mathsf{y} = 0$ and $(\mathsf{x} + \mathsf{y}) \cdot (\mathsf{x} + \mathsf{y}) = 0$ without having $\mathsf{x} \cdot \mathsf{y} = 0$. However, this criterion may result in a self-orthogonal code. For instance, the first code in Table 1 is self-orthogonal, whereas the second code is not.

**Lemma 5** *The property $P[\mathsf{x}]$ is true if and only if $w_E(\mathsf{x}) \equiv 0 \mod 8$ is a multiplicative property over $\mathbb{Z}_4^n$.*

**Proof.** Let $\mathsf{x} \in \mathbb{Z}_4^n$ such that $w_E(\mathsf{x}) = n_1(\mathsf{x}) + 4n_2(\mathsf{x}) + n_3(\mathsf{x}) \equiv 0 \mod 8$. We must prove that $w_E(3\mathsf{x}) \equiv 0 \mod 8$. We have $w_E(\mathsf{x}) = w_E(3\mathsf{x})$, because $n_1(3\mathsf{x}) = n_3(\mathsf{x})$, $n_3(3\mathsf{x}) = n_1(\mathsf{x})$ and $n_2(3\mathsf{x}) = n_2(\mathsf{x})$. Since we have assumed

Table 1: Lexicodes over $\mathbb{Z}_4^n$ Obtained using the Selection Property $\mathsf{x} \cdot \mathsf{x} = 0$

| $n$ | Basis of $\mathbb{Z}_4^n$ | Basis of $C(B,P)$ | Type | $d_L$ |
|---|---|---|---|---|
| 4 | Canonical basis | $\mathsf{a}_1 = 2000$<br>$\mathsf{a}_2 = 0200$<br>$\mathsf{a}_3 = 0020$<br>$\mathsf{a}_4 = 1111$ | $42^3$ | 2 |
| 4 | $\mathsf{b}_1 = 0001$<br>$\mathsf{b}_2 = 1100$<br>$\mathsf{b}_3 = 0110$<br>$\mathsf{b}_4 = 0011$ | $\mathsf{a}_1 = 0002$<br>$\mathsf{a}_2 = 2200$<br>$\mathsf{a}_3 = 0220$<br>$\mathsf{a}_4 = 1111$ | $42^3$ | 2 |
| 6 | Canonical basis | $\mathsf{a}_1 = 200000$<br>$\mathsf{a}_2 = 020000$<br>$\mathsf{a}_3 = 002000$<br>$\mathsf{a}_4 = 111100$<br>$\mathsf{a}_5 = 000020$<br>$\mathsf{a}_6 = 110011$ | $4^2 2^4$ | 2 |
| 6 | $\mathsf{b}_1 = 322323$<br>$\mathsf{b}_2 = 220033$<br>$\mathsf{b}_3 = 311201$<br>$\mathsf{b}_4 = 322122$<br>$\mathsf{b}_5 = 212130$<br>$\mathsf{b}_6 = 231230$ | $\mathsf{a}_1 = 200202$<br>$\mathsf{a}_2 = 000022$<br>$\mathsf{a}_3 = 311201$<br>$\mathsf{a}_4 = 102111$<br>$\mathsf{a}_5 = 0202220$<br>$\mathsf{a}_6 = 022020$ | $4^5 2$ | 2 |
| 8 | Canonical basis | $\mathsf{a}_1 = 20000000$<br>$\mathsf{a}_2 = 02000000$<br>$\mathsf{a}_3 = 00200000$<br>$\mathsf{a}_4 = 11110000$<br>$\mathsf{a}_5 = 00002000$<br>$\mathsf{a}_6 = 11001100$<br>$\mathsf{a}_7 = 10000001$<br>$\mathsf{a}_8 = 01101001$ | $4^4 2^4$ | 2 |

that $w_E(\mathsf{x}) \equiv 0 \mod 8$, the result follows. ■

**Remark 2** *The condition given in Lemma 5 is a sufficient condition to obtain self-orthogonal codes over $\mathbb{Z}_4$ [19, Theorem 12.2.4]. Hence by applying Algorithm A with the property $P[\mathsf{x}] = w_E(\mathsf{x}) \equiv 0 \mod 8$, we obtain self-orthogonal codes.*

Self-orthogonal lexicodes over $\mathbb{Z}_4^n$ obtained using the selection property $w_E(\mathsf{x}) \equiv 0 \mod 8$ are given in Table 2. Note that some of these codes are self-dual. The minimum Lee distance of the codes are compared with those given in [2] and [12]. The symbol $\Diamond$ denotes that there is no result to compare with, and $\times$ denotes that the code is not self-dual.

## 4.2 Lexicodes with a Weight Criteria

For $\delta$ a positive integer, from Lemma 2 the property $w_L(\mathsf{x}) \geq \delta$ is multiplicative. Therefore we have the following result.

**Corollary 6** *The lexicode $C(B, \delta)$ given by Algorithm A for the selection property $P[\mathsf{x}]$ if and only if $w_L(\mathsf{x}) \geq \delta$ is a linear code over $\mathbb{Z}_4$ with minimum Lee distance greater than or equal to $\delta$.*

Several lexicodes over $\mathbb{Z}_4^n$ obtained using the selection property $w_L(\mathsf{x}) \geq \delta$ are given in Tables 3 and 4.

**Remark 3** *The selection property on the Lee weight gives codes with good parameters. For example, the seventh code in Table 4 is the self-dual octacode. In the next section we will prove that their binary images are also good.*

Table 2: Self-orthogonal Lexicodes over $\mathbb{Z}_4^n$ Obtained using the Selection Property $w_E(\mathsf{x}) \equiv 0 \mod 8$

| $n$ | Basis of $\mathbb{Z}_4^n$ | Basis of $C(B,P)$ | Type | $d_L$ | [2] | [12] | Self-dual |
|---|---|---|---|---|---|---|---|
| 4 | $\mathsf{b}_1 = 0001$<br>$\mathsf{b}_2 = 1100$<br>$\mathsf{b}_3 = 0110$<br>$\mathsf{b}_4 = 0011$ | $\mathsf{a}_1 = 2200$<br>$\mathsf{a}_2 = 0220$<br>$\mathsf{a}_3 = 0022$ | $2^3$ | 4 | $\Diamond$ | 4 | $\times$ |
| 5 | $\mathsf{b}_1 = 01010$<br>$\mathsf{b}_2 = 10100$<br>$\mathsf{b}_3 = 33100$<br>$\mathsf{b}_4 = 00003$<br>$\mathsf{b}_5 = 00100$ | $\mathsf{a}_1 = 02020$<br>$\mathsf{a}_2 = 20200$<br>$\mathsf{a}_3 = 22000$<br>$\mathsf{a}_4 = 11112$ | $2^3 4$ | 4 | 4 | 4 | $\times$ |
| 6 | Canonical basis | $\mathsf{a}_1 = 220000$<br>$\mathsf{a}_2 = 202000$<br>$\mathsf{a}_3 = 200200$<br>$\mathsf{a}_4 = 200020$<br>$\mathsf{a}_5 = 200002$ | $2^5$ | 4 | $\Diamond$ | 4 | $\times$ |
| 6 | $\mathsf{b}_1 = 322323$<br>$\mathsf{b}_2 = 220033$<br>$\mathsf{b}_3 = 311201$<br>$\mathsf{b}_4 = 322122$<br>$\mathsf{b}_5 = 212130$<br>$\mathsf{b}_6 = 231230$ | $\mathsf{a}_1 = 000022$<br>$\mathsf{a}_2 = 222002$<br>$\mathsf{a}_3 = 102111$<br>$\mathsf{a}_4 = 222222$ | $4^1 2^3$ | 4 | $\Diamond$ | 4 | $\times$ |
| 8 | $\mathsf{b}_1 = 32121211$<br>$\mathsf{b}_2 = 01132301$<br>$\mathsf{b}_3 = 23002111$<br>$\mathsf{b}_4 = 22231202$<br>$\mathsf{b}_5 = 11200323$<br>$\mathsf{b}_6 = 01312220$<br>$\mathsf{b}_7 = 20121213$<br>$\mathsf{b}_8 = 31012112$ | $\mathsf{a}_1 = 22022220$<br>$\mathsf{a}_2 = 02000222$<br>$\mathsf{a}_3 = 00022000$<br>$\mathsf{a}_4 = 22000202$<br>$\mathsf{a}_5 = 22022022$<br>$\mathsf{a}_6 = 00202022$<br>$\mathsf{a}_7 = 13331313$ | $4^1 2^6$ | 4 | $\Diamond$ | $\Diamond$ | Self-dual |
| 8 | $\mathsf{b}_1 = 11112233$<br>$\mathsf{b}_2 = 23100323$<br>$\mathsf{b}_3 = 02222133$<br>$\mathsf{b}_4 = 01133231$<br>$\mathsf{b}_5 = 21310130$<br>$\mathsf{b}_6 = 23101130$<br>$\mathsf{b}_7 = 23001233$<br>$\mathsf{b}_8 = 11203211$ | $\mathsf{a}_1 = 22220022$<br>$\mathsf{a}_2 = 02200202$<br>$\mathsf{a}_3 = 02222022$<br>$\mathsf{a}_4 = 02220002$<br>$\mathsf{a}_5 = 11131331$<br>$\mathsf{a}_6 = 02002022$<br>$\mathsf{a}_7 = 22002200$ | $4^1 2^6$ | 4 | $\Diamond$ | $\Diamond$ | Self-dual |
| 9 | $\mathsf{b}_1 = 121221011$<br>$\mathsf{b}_2 = 232312211$<br>$\mathsf{b}_3 = 010102101$<br>$\mathsf{b}_4 = 131023121$<br>$\mathsf{b}_5 = 233011332$<br>$\mathsf{b}_6 = 300221122$<br>$\mathsf{b}_7 = 103131120$<br>$\mathsf{b}_8 = 222032231$<br>$\mathsf{b}_9 = 210312111$ | $\mathsf{a}_1 = 222222000$<br>$\mathsf{a}_2 = 010102101$<br>$\mathsf{a}_3 = 320102312$<br>$\mathsf{a}_4 = 002000222$<br>$\mathsf{a}_5 = 000200200$<br>$\mathsf{a}_6 = 222002022$ | $4^2 2^4$ | 4 | $\Diamond$ | $\Diamond$ | $\times$ |

Table 3: Lexicodes over $\mathbb{Z}_4^n$ Obtained using the Selection Property $w_L(\mathsf{x}) \geq \delta$

| $n$ | Basis of $\mathbb{Z}_4^n$ | $\delta$ | Basis of $C(B,P)$ | Type | $d_L$ |
|---|---|---|---|---|---|
| 3 | Canonical basis | 2 | $\mathsf{a}_1 = 110$<br>$\mathsf{a}_2 = 101$ | $4^2 2$ | 2 |
| 4 | $\mathsf{b}_1 = 0001$<br>$\mathsf{b}_2 = 1100$<br>$\mathsf{b}_3 = 0110$<br>$\mathsf{b}_4 = 0011$ | 2 | $\mathsf{a}_1 = 1100$<br>$\mathsf{a}_2 = 0110$<br>$\mathsf{a}_3 = 0011$ | $4^3 2$ | 2 |
| 5 | Canonical basis | 3 | $\mathsf{a}_1 = 11100$<br>$\mathsf{a}_2 = 21010$<br>$\mathsf{a}_3 = 31001$ | $4^3$ | 3 |
| 5 | $\mathsf{b}_1 = 10100$<br>$\mathsf{b}_2 = 10010$<br>$\mathsf{b}_3 = 33100$<br>$\mathsf{b}_4 = 00003$<br>$\mathsf{b}_5 = 00100$ | 3 | $\mathsf{a}_1 = 11110$<br>$\mathsf{a}_2 = 33103$ | $4^2$ | 3 |
| 6 | Canonical basis | 4 | $\mathsf{a}_1 = 211000$<br>$\mathsf{a}_2 = 12011$<br>$\mathsf{a}_3 = 200011$ | $4^3$ | 4 |
| 6 | $\mathsf{b}_1 = 231311$<br>$\mathsf{b}_2 = 122322$<br>$\mathsf{b}_3 = 122101$<br>$\mathsf{b}_4 = 211321$<br>$\mathsf{b}_5 = 110321$<br>$\mathsf{b}_6 = 132023$ | 2 | $\mathsf{a}_1 = 231311$<br>$\mathsf{a}_2 = 122322$<br>$\mathsf{a}_2 = 122101$<br>$\mathsf{a}_2 = 312221$ | $4^4$ | 2 |
| 6 | ” | 3 | $\mathsf{a}_1 = 231311$<br>$\mathsf{a}_2 = 122101$<br>$\mathsf{a}_3 = 333203$ | $4^3$ | 4 |
| 6 | ” | 4 | $\mathsf{a}_1 = 231311$<br>$\mathsf{a}_2 = 122101$<br>$\mathsf{a}_3 = 210001$ | $4^3$ | 4 |
| 6 | ” | 5 | $\mathsf{a}_1 = 231311$<br>$\mathsf{a}_2 = 122101$ | $4^2$ | 5 |
| 6 | ” | 6 | $\mathsf{a}_1 = 231311$ | 4 | 7 |
| 8 | $\mathsf{b}_1 = 22312221$<br>$\mathsf{b}_2 = 11311303$<br>$\mathsf{b}_3 = 00121200$<br>$\mathsf{b}_4 = 01313032$<br>$\mathsf{b}_5 = 30122132$<br>$\mathsf{b}_6 = 03213232$<br>$\mathsf{b}_7 = 32132232$<br>$\mathsf{b}_8 = 12201321$ | 5 | $\mathsf{a}_1 = 22312221$<br>$\mathsf{a}_2 = 11311303$<br>$\mathsf{a}_3 = 01030232$ | $4^3$ | 5 |
| 8 | $\mathsf{b}_1 = 11112233$<br>$\mathsf{b}_2 = 23100323$<br>$\mathsf{b}_3 = 02222133$<br>$\mathsf{b}_4 = 01133231$<br>$\mathsf{b}_5 = 21310130$<br>$\mathsf{b}_6 = 23101130$<br>$\mathsf{b}_7 = 23001233$<br>$\mathsf{b}_8 = 11203211$ | 2 | $\mathsf{a}_1 = 11112233$<br>$\mathsf{a}_2 = 23100323$<br>$\mathsf{a}_3 = 02222133$<br>$\mathsf{a}_4 = 01133231$<br>$\mathsf{a}_5 = 21310130$<br>$\mathsf{a}_6 = 20311130$<br>$\mathsf{a}_7 = 22301233$ | $4^7$ | 2 |

Table 4: Lexicodes over $\mathbb{Z}_4^n$ Obtained using the Selection Property $w_L(\mathsf{x}) \geq \delta$

| $n$ | Basis of $\mathbb{Z}_4^n$ | $\delta$ | Basis of $C(B,P)$ | Type | $d_L$ |
|---|---|---|---|---|---|
| 8 | | 3 | $\mathsf{a}_1 = 11112233$<br>$\mathsf{a}_2 = 23100323$<br>$\mathsf{a}_3 = 02222133$<br>$\mathsf{a}_4 = 21310130$<br>$\mathsf{a}_5 = 22133112$ | $4^5$ | 3 |
| 8 | " | 4 | $\mathsf{a}_1 = 11112233$<br>$\mathsf{a}_2 = 23100323$<br>$\mathsf{a}_3 = 02222133$<br>$\mathsf{a}_4 = 23132112$ | $4^4$ | 4 |
| 8 | | 5 | $\mathsf{a}_1 = 11112233$<br>$\mathsf{a}_2 = 23100323$<br>$\mathsf{a}_3 = 02222133$ | $4^3$ | 5 |
| 8 | | 6 | $\mathsf{a}_1 = 11112233$<br>$\mathsf{a}_2 = 23100323$<br>$\mathsf{a}_3 = 33033123$ | $4^3$ | 6 |
| 8 | | 7 | $\mathsf{a}_1 = 11112233$ | $4^1$ | 10 |
| 8 | $\mathsf{b}_1 = 10003121$<br>$\mathsf{b}_2 = 01001231$<br>$\mathsf{b}_3 = 00103332$<br>$\mathsf{b}_4 = 00012311$<br>$\mathsf{b}_5 = 22233221$<br>$\mathsf{b}_6 = 10302221$<br>$\mathsf{b}_7 = 10312111$<br>$\mathsf{b}_8 = 02311100$ | 2 | $\mathsf{a}_1 = 10003121$<br>$\mathsf{a}_2 = 01001231$<br>$\mathsf{a}_3 = 00103332$<br>$\mathsf{a}_4 = 00012311$<br>$\mathsf{a}_5 = 22233221$<br>$\mathsf{a}_6 = 10302221$ | $4^6$ | 2 |
| 8 | | $3 \leq \delta \leq 6$ | $\mathsf{a}_1 = 10003121$<br>$\mathsf{a}_2 = 01001231$<br>$\mathsf{a}_3 = 00103332$<br>$\mathsf{a}_4 = 00012311$ | $4^4$ | 6 |
| 8 | | 7 | $\mathsf{a}_1 = 21102321$<br>$\mathsf{a}_2 = 10310132$ | $4^2$ | 7 |
| 8 | | 8 | $\mathsf{a}_1 = 21102321$<br>$\mathsf{a}_2 = 21213100$ | $4^2$ | 8 |
| 9 | $\mathsf{b}_1 = 121221011$<br>$\mathsf{b}_2 = 232312211$<br>$\mathsf{b}_3 = 010102101$<br>$\mathsf{b}_4 = 131023121$<br>$\mathsf{b}_5 = 233011332$<br>$\mathsf{b}_6 = 300221122$<br>$\mathsf{b}_7 = 103131120$<br>$\mathsf{b}_8 = 222032231$<br>$\mathsf{b}_9 = 210312111$ | 8 | $\mathsf{a}_1 = 121221011$<br>$\mathsf{a}_2 = 323311112$ | $4^2$ | 8 |
| 10 | $\mathsf{b}_1 = 2212122203$<br>$\mathsf{b}_2 = 0123002220$<br>$\mathsf{b}_3 = 0023010100$<br>$\mathsf{b}_4 = 1010312112$<br>$\mathsf{b}_5 = 2111023221$<br>$\mathsf{b}_6 = 1211332321$<br>$\mathsf{b}_7 = 3110131311$<br>$\mathsf{b}_8 = 0313130000$<br>$\mathsf{b}_9 = 1202313120$<br>$\mathsf{b}_{10} = 1122001000$ | 8 | $\mathsf{a}_1 = 2331120023$<br>$\mathsf{a}_2 = 0302111120$<br>$\mathsf{a}_3 = 3001103202$ | $4^3$ | 8 |

## 4.3 Good Binary Codes from Lexicodes over $\mathbb{Z}_4$

It was shown by Hammons et al. [17] that some of the best known nonlinear binary codes such as the Nordstrom-Robinson, Kerdock, Preparata, Goethals and Delsarte-Goethals codes are Gray map images of $\mathbb{Z}_4$-linear codes. The Gray map from $\mathbb{Z}_4$ to $\mathbb{F}_2^2$ is defined as

$$\mathcal{G}'(0) = 00, \mathcal{G}'(1) = 01, \mathcal{G}'(2) = 11, \mathcal{G}'(3) = 10.$$

The Gray map $\mathcal{G} : \mathbb{Z}_4^n \longrightarrow \mathbb{F}_2^{2n}$ is then defined as

$$\mathcal{G}(a_1, \dots a_n) = (\mathcal{G}'(a_1), \dots, \mathcal{G}'(a_n)).$$

The following result is well known.

**Lemma 7** *The Gray map $\mathcal{G}$ is the distance-preserving map*

$$(\mathbb{Z}_4^n,\ \textit{Lee distance}) \longrightarrow (\mathbb{F}_2^{2n},\ \textit{Hamming distance}).$$

The covering radius of a code $C$ over $\mathbb{Z}_4$ with respect to the Lee distance is defined as

$$\rho_L(C) = \max_{u \in \mathbb{Z}_4^n} \{\min_{c \in C} d_L(u, c)\}.$$

For $u \in \mathbb{Z}_4^n$, the coset of $C$ is defined to be the set $u + C = \{u + c | c \in C\}$. A minimum weight vector in a coset is called a coset leader. It is obvious that the covering radius of $C$ with respect to the Lee weight is the largest minimum weight among all cosets.

**Lemma 8** *( [1, Proposition 3.2]) Let $C$ be a code over $\mathbb{Z}_4$ with $\mathcal{G}(C)$ the Gray map image of $C$. Then*

$$\rho_L(C) = \rho(\mathcal{G}(C)).$$

**Proposition 9** *Let $0 = C_0 \subseteq C_1 \subseteq \cdots \subseteq C_n = C(B, \delta)$ be the set of nested codes obtained by Algorithm A for designed distance $\delta$. Hence for $0 \leq i < n$ if the $C_i$, are of type $4^{k_{i_1}} 2^{k_{i_2}}$, the covering radius $\rho_L(C_i)$ satisfies*

$$\delta \leq \rho_L(C_i) \leq 2(n - k_{i_1}) - k_{i_2}. \tag{2}$$

*Then we have*

$$\lfloor \delta/2 \rfloor \leq \lfloor d/2 \rfloor \leq \rho_L(C(B, \delta)) \leq \delta - 1 \leq d - 1. \tag{3}$$

**Proof.** If for all $i \leq n$, we have $C_0 = C_1 = \ldots = C_n$, then the lexicode is trivial. Hence assume that $C_i \subsetneq C_n$ for some $0 \leq i < n$. Now, let $x \in C_n \setminus C_i$ be a codeword of minimum weight. Such a vector must be a coset leader of $C_i$, as $C_i \subsetneq C_n$. Hence $\rho_L(C_i) \geq wt_L(\mathsf{x})$ and then $\rho_L(C_i) \geq \delta$. The right side of (2) is obtained from the redundancy bound [1, Theorem 4.6]. Since each vector in $\mathbb{Z}_4^n$ has distance $\delta - 1$ or less to some vector in $C_n$, the covering radius of $C_n$ is at most $\delta - 1$. By the construction we have $\lfloor \delta/2 \rfloor \leq \lfloor d/2 \rfloor$. The left side of (3) is obtained from the packing radius bound [1, Theorem

4.3]. ■

**Theorem 10** *Let $C_L(B,\delta)$ be the lexicode obtained by Algorithm A. Then the binary code $\mathcal{G}(C_L(B,\delta))$ obtained from $C_L(B,\delta)$ by the Gray image meets the Gilbert bound.*

**Proof.** Assume that $\mathcal{G}(C_n)$ is a binary code of minimum distance $d$, which is the same as the minimum distance of $C_L(B,\delta)$ since the Gray map is a weight preserving map. Hence we have $d \geq \delta$, and by Lemma 8 $\rho(\mathcal{G}(C_L(B,\delta)) = \rho_L(C_L(B,\delta))$. Then from Proposition 9 we have $\rho_L(C_L(B,\delta)) \leq \delta - 1$. Since $\delta \leq d$, $\mathcal{G}(C_n)$ has covering radius less than $d_L - 1$. It is well known [19, p. 87], that a code over $\mathbb{F}_q$ with minimum distance $d$ and covering radius $d-1$ or less meets the Gilbert bound. ■

# 5 Construction of Lexicodes over $\mathbb{F}_2 + u\mathbb{F}_2$

In this section, for simplicity we denote the ring $\mathbb{F}_2 + u\mathbb{F}_2$ by $R$. $R$ is a finite chain ring with 4 elements, maximal ideal $\langle u \rangle$ and nilpotency index 2. The set of units of $R$ is $R^* = \{1, \overline{u}\}$. There is a Gray map $\Phi$ that is an $\mathbb{F}_2$−linear isometry from ($R^n$, Lee distance) to ($\mathbb{F}_2^{2n}$, Hamming distance), and is given by $\Phi(x+uy) = (y, x+y)$. An interesting fact regarding the Gray map over $R$ is that the image of a self-dual linear code $C$ over $R$ is a self-dual linear binary code [11].

Assume now that we have a property $P$ which can test if a vector $\mathsf{c} \in R^n$

is selected or not. Hence $P$ is multiplicative if for $P[\mathsf{x}]$ then $P[\overline{u}\mathsf{x}]$, since $R^* = \{1, \overline{u}\}$. We now prove the following result.

**Proposition 11** *Let $C$ be a self-orthogonal code over $R$. Then for all $\mathsf{x} \in C$ we have $w_L(\mathsf{x}) \equiv 0 \mod 2$.*

**Proof.** Let $C$ be a self-orthogonal code and $\mathsf{x}$ be a codeword of $C$. Then we have $\mathsf{x} \cdot \mathsf{x} = 0 \mod 2$, but

$$\mathsf{x} \cdot \mathsf{x} = \sum_{1}^{n} x_i x_i = \sum_{1}^{n_1(\mathsf{x})} 1 + \sum_{1}^{n_u(\mathsf{x})} u^2 + \sum_{1}^{n_{\overline{u}}(\mathsf{x})} \overline{u}^2 = n_1(\mathsf{x}) + n_{\overline{u}}(\mathsf{x}) \equiv 0 \mod 2.$$

This gives $w_L(\mathsf{x}) \equiv 0 \mod 2$. ■

Since the Lee weight is a homogeneous weight over $R$, from Definition 1 it is obvious that the property $P[\mathsf{x}]$ is true if and only if $w_L(\mathsf{x}) \equiv 0 \mod 2$ is a multiplicative property. Using Algorithm A over $R$ with this selection property, we obtained the codes given in Table 5.

From Lemma 3 and Theorem 4 we have the following.

**Corollary 12** *The lexicode $C(B, \delta)$ given by Algorithm A over $R$ for the selection property $P[\mathsf{x}]$ if and only if $w_L(\mathsf{x}) \geq \delta$ is a linear code over $R$ with Lee minimum distance greater than or equal to $\delta$.*

**Remark 4** *With few exceptions, the binary images of the codes given in Table 6 are optimal codes according to [15].*

Table 5: Lexicodes over $R^n$ Obtained using the Selection Property $w_L(\mathsf{x}) \equiv 0$ mod 2

| $n$ | Basis of $R^n$ | Basis of $C(B,P)$ | Type | $d_L$ |
|---|---|---|---|---|
| 4 | Canonical basis | $\mathsf{a}_1 = u000$<br>$\mathsf{a}_2 = 1010$<br>$\mathsf{a}_3 = 1100$<br>$\mathsf{a}_4 = 1001$ | $4^3 2$ | 2 |
| 4 | $\mathsf{b}_1 = 1100$<br>$\mathsf{b}_2 = 1u01$<br>$\mathsf{b}_3 = \overline{u}11\overline{u}$<br>$\mathsf{b}_4 = \overline{u}\overline{u}0\overline{u}$ | $\mathsf{a}_1 = 1100$<br>$\mathsf{a}_2 = 1u01$<br>$\mathsf{a}_3 = \overline{u}11\overline{u}$<br>$\mathsf{a}_4 = uu0u$ | $4^3 2$ | 2 |
| 6 | $\mathsf{b}_1 = 1u\overline{u}0\overline{u}u$<br>$\mathsf{b}_2 = \overline{u}u\overline{u}uu1$<br>$\mathsf{b}_3 = 0\overline{u}\overline{u}u11$<br>$\mathsf{b}_4 = u01u1u$<br>$\mathsf{b}_5 = 1uu101$<br>$\mathsf{b}_6 = 1\overline{u}01u\overline{u}$ | $\mathsf{a}_1 = u0u0u0$<br>$\mathsf{a}_2 = u00u1\overline{u}$<br>$\mathsf{a}_3 = 0\overline{u}\overline{u}u11$<br>$\mathsf{a}_4 = u01u1u$<br>$\mathsf{a}_5 = 0011\overline{u}\overline{u}$<br>$\mathsf{a}_6 = 1\overline{u}01u\overline{u}$ | $4^5 2$ | 2 |
| 6 | $\mathsf{b}_1 = \overline{u}00u11$<br>$\mathsf{b}_2 = uu0\overline{u}u\overline{u}$<br>$\mathsf{b}_3 = 101110$<br>$\mathsf{b}_4 = 110001$<br>$\mathsf{b}_5 = u01011$<br>$\mathsf{b}_6 = u10uuu$ | $\mathsf{a}_1 = u000uu$<br>$\mathsf{a}_2 = uu0\overline{u}u\overline{u}$<br>$\mathsf{a}_3 = 101110$<br>$\mathsf{a}_4 = u10u10$<br>$\mathsf{a}_5 = u10u10$<br>$\mathsf{a}_6 = 1100\overline{u}\overline{u}$ | $4^5 2$ | 2 |
| 6 | $\mathsf{b}_1 = 0u0\overline{u}0u$<br>$\mathsf{b}_2 = 0\overline{u}0\overline{u}10$<br>$\mathsf{b}_3 = 10\overline{u}100$<br>$\mathsf{b}_4 = 1\overline{u}110u$<br>$\mathsf{b}_5 = 001100$<br>$\mathsf{b}_6 = 00\overline{u}\overline{u}11$ | $\mathsf{a}_1 = 000u00$<br>$\mathsf{a}_2 = 01001u$<br>$\mathsf{a}_3 = 1u\overline{u}u0u$<br>$\mathsf{a}_4 = 1\overline{u}110u$<br>$\mathsf{a}_5 = 001100$<br>$\mathsf{a}_6 = 00\overline{u}\overline{u}11$ | $4^5 2$ | 2 |
| 8 | $\mathsf{b}_1 = \overline{u}1111uuu$<br>$\mathsf{b}_2 = 0\overline{u}00u01\overline{u}$<br>$\mathsf{b}_3 = 000uuu11$<br>$\mathsf{b}_4 = u\overline{u}u1\overline{u}111$<br>$\mathsf{b}_5 = u\overline{u}u1\overline{u}110$<br>$\mathsf{b}_6 = 100\overline{u}1\overline{u}u1$<br>$\mathsf{b}_7 = uu\overline{u}\overline{u}11\overline{u}1$<br>$\mathsf{b}_8 = 0\overline{u}u1u1uu$ | $\mathsf{a}_1 = uuuuu000$<br>$\mathsf{a}_2 = \overline{u}u11\overline{u}u\overline{u}1$<br>$\mathsf{a}_3 = \overline{u}11uu1\overline{u}\overline{u}$<br>$\mathsf{a}_4 = u\overline{u}u1\overline{u}111$<br>$\mathsf{a}_5 = 1u\overline{u}0u\overline{u}\overline{u}u$<br>$\mathsf{a}_6 = u11u010\overline{u}$<br>$\mathsf{a}_5 = uu\overline{u}\overline{u}11\overline{u}1$<br>$\mathsf{a}_6 = \overline{u}u\overline{u}0\overline{u}\overline{u}00$ | $4^7 2$ | 2 |

Table 6: Lexicodes over $R^n$ Obtained using the Selection Property $w_L(\mathsf{x}) \geq \delta$

| $n$ | Basis of $R^n$ | $\delta$ | Basis of $C(B, P)$ | Type | $d_L$ |
|---|---|---|---|---|---|
| 6 | Canonical basis | 4 | $\mathsf{a}_1 = u11000$<br>$\mathsf{a}_2 = 1u0100$<br>$\mathsf{a}_3 = u00011$ | $4^3$ | 4 |
| 6 | $\mathsf{b}_1 = 1u\overline{u}0\overline{u}u$<br>$\mathsf{b}_2 = \overline{u}u\overline{u}uu1$<br>$\mathsf{b}_3 = 0\overline{u}\overline{u}u11$<br>$\mathsf{b}_4 = u01u1u$<br>$\mathsf{b}_5 = 1uu101$<br>$\mathsf{b}_6 = 1\overline{u}01u\overline{u}$ | 4 | $\mathsf{a}_1 = 1u\overline{u}0\overline{u}u$<br>$\mathsf{a}_2 = \overline{u}u\overline{u}uu1$ | $4^2$ | 4 |
| 6 | $\mathsf{b}_1 = 0u0\overline{u}0u$<br>$\mathsf{b}_2 = 0\overline{u}0\overline{u}10$<br>$\mathsf{b}_3 = 10\overline{u}100$<br>$\mathsf{b}_4 = 1\overline{u}110u$<br>$\mathsf{b}_5 = 001\overline{u}00$<br>$\mathsf{b}_6 = 00\overline{u}\overline{u}11$ | 5 | $a_1 = 11\overline{u}\overline{u}1u$<br>$a_2 = 0\overline{u}\overline{u}uu1$ | $4^2$ | 5 |
| 6 | $\mathsf{b}_1 = 0u0\overline{u}0u$<br>$\mathsf{b}_2 = 0\overline{u}0\overline{u}10$<br>$\mathsf{b}_3 = 10\overline{u}100$<br>$\mathsf{b}_4 = 1\overline{u}110u$<br>$\mathsf{b}_5 = 001\overline{u}00$<br>$\mathsf{b}_6 = 00\overline{u}\overline{u}11$ | 4 | $\mathsf{a}_1 = 010u1u$<br>$\mathsf{a}_2 = 1u\overline{u}00u$<br>$\mathsf{a}_3 = 1\overline{u}110u$ | $4^3$ | 4 |
| 6 | $\mathsf{b}_1 = 0000u1$<br>$\mathsf{b}_2 = \overline{u}00\overline{u}11$<br>$\mathsf{b}_3 = \overline{u}u1100$<br>$\mathsf{b}_4 = u\overline{u}u1\overline{u}0$<br>$\mathsf{b}_5 = \overline{u}1u\overline{u}u1$<br>$\mathsf{b}_6 = u\overline{u}u10u$ | 5 | $\mathsf{a}_1 = \overline{u}00\overline{u}\overline{u}u$<br>$\mathsf{a}_2 = \overline{u}u11u1$ | $4^2$ | 5 |
| 6 | $\mathsf{b}_1 = 0000u1$<br>$\mathsf{b}_2 = \overline{u}00\overline{u}11$<br>$\mathsf{b}_3 = \overline{u}u1100$<br>$\mathsf{b}_4 = u\overline{u}u1\overline{u}0$<br>$\mathsf{b}_5 = \overline{u}1u\overline{u}u1$<br>$\mathsf{b}_6 = u\overline{u}u10u$ | 4 | $\mathsf{a}_1 = \overline{u}00\overline{u}11$<br>$\mathsf{a}_2 = \overline{u}u1100$<br>$\mathsf{a}_3 = \overline{u}1u\overline{u}u1$ | $4^3$ | 4 |
| 8 | $\mathsf{b}_1 = \overline{u}u1u1u11$<br>$\mathsf{b}_2 = 011\overline{u}u\overline{u}01$<br>$\mathsf{b}_3 = u\overline{u}00u111$<br>$\mathsf{b}_4 = uuu\overline{u}1u0u$<br>$\mathsf{b}_5 = 11u00\overline{u}u\overline{u}$<br>$\mathsf{b}_6 = 01\overline{u}1uuu0$<br>$\mathsf{b}_7 = u01u1u1\overline{u}$<br>$\mathsf{b}b_8 = \overline{u}101u11u$ | 5 | $a_1 = \overline{u}u1u1u11$<br>$a_2 = 011\overline{u}u\overline{u}01$<br>$a_3 = 111u\overline{u}\overline{u}uu$<br>$a_4 = \overline{u}\overline{u}1uuu\overline{u}u$ | $4^4$ | 5 |

**Acknowledgements**

The authors would like to thank reviewers for useful comments which improved the paper considerably. As well as the Editor for wonderful job that has been done.